# Experimental investigation of perforations interactions effects under high sound pressure levels

**Rostand Tayong and Philippe Leclaire**

Laboratoire de Recherche en Mécanique et Acoustique
Université de Bourgogne, Nevers, France

Most models for predicting the acoustic response of perforated panels are based on the assumption that there is no interaction between the holes. The common way of taking into account the perforates effects is the use of Fok's functions. The few papers that deal with these effects study the case of low sound pressure levels. This paper investigates the Holes Interaction Effects (HIE) of perforated and micro-perforated panels when submitted to high sound pressure levels. Analysis of the data yields to the fact that even with HIE, the nonlinear reactance dependency with velocity is very slight. The HIE can provide good absorption of the perforated panel backed by an air cavity at low and high sound pressure levels if the holes positions are well configured. Perforated panels of holes diameters less than 2 millimeters were built out of steel with different interstices (distance between two adjacent holes) and tested. Experimental results (performed with an impedance tube) in the cases of interaction are done and compared with the exact cases of no interactions. The results can be used for designing optimal perforated panels for ducts silencers for instance.

## Introduction

In various noise control applications such as ducts, exhaust systems and aircraft, perforated panels are used to attenuate sound. One of the advantages of such acoustical materials is that their frequency resonances can be tuned depending on the goal to achieve. When the perforations are reduced to millimeter or sub-millimeter (micro-perforation) size, these materials can afford very interesting sound absorption without any additional classical absorbing material. Besides, they are proved to be very useful in dealing with the low-frequency noise.

A great number of models were proposed to modeling the acoustical behavior of such systems. The particularity of most of these models [1, 2, 3] is that they are only applicable for widely separated holes (assumption of no interaction between the perforations). Ingard [4] did an extensive survey on the topic of resonators. In his work, he considered the case of two apertures interacting and came up with the fact that the end-correction is very dependent on the holes separation. The common study dealing with the holes interaction effect may be related to Fok [5]. From his work [5, 6], a function (called the Fok function) was derived taking into account the distance between apertures and the correction for the radiating impedance of interacting perforations. Later on, Melling [7] reconsidered this function and noticed that this correction of Fok will be in practice small for low porosity samples but for high porosity samples will reduce the end-correction by a significant amount. Moreover, Randeberg [8] particularly revealed that the result of applying Fok function as a correction due to the hole correction is practically equivalent to using Ingard's inner end correction for both apertures of holes in a perforated panel. If the latter studies were focused on the case of low sound intensities, there is no work known to us presently, dealing specifically with holes interactions effects under high sound intensities.

The objective of this paper is to investigate the holes interaction effects on the sound absorption of perforated panels under high sound pressure levels. In a first step, an aspect of the open area ratio of perforated panels is presented, followed by a model based on Ingard's theory for acoustic resonators accounting of the interactions effects. Secondly, the experimental setup is described. Then, the Measurements results for the cases of interactions are used for the analysis and compared to the exact cases of no interactions. The important results are summarized in the conclusion of this paper.

# 1. The theory for holes interaction effect

## 1.1 The open area ratio

The open area ratio *σ* is defined as the ratio of the total area of the holes to the total area $S_p$ of the panel. Assuming cylindrical perforations with the same diameter *a*, the open area ratio is given by:

$$\sigma = \frac{n\pi a^2}{4S_p}. \quad (1)$$

where *n* is the number of holes on the panel. Two cases may be noticed depending on the fact that the holes are evenly distributed or not. If the holes are evenly distributed on the total panel area, there is no need of knowing the number of holes on the panel. The diameter *a* of holes and the distance *b* between two consecutive holes are sufficient to determine *σ*. The formula is then:

$$\sigma = \frac{na^2}{4b^2}. \quad (2)$$

This is probably the most used formula to calculate *σ*. Now, instead of using *b* (the distance between 2 consecutive holes centers) one may use the interstice length β, a more interesting distance, which is the distance between 2 holes edges (see Fig. 1). β is simply given by :

$$\beta = b - a. \quad (3)$$

Equation (2) becomes :

$$\sigma = \frac{na^2}{4(\beta + a)^2}. \quad (4)$$

In the case where the holes are not evenly distributed on the panel total area as on figure 1 (holes localized in a part of the panel area for instance), assuming regular distance between the holes, *σ* is given by :

$$\sigma = \frac{n\pi a^2}{4(nb^2 + S_u)}, \quad (5)$$

where $S_u$ is the remaining area of the panel that would be pierced if the holes were evenly distributed on the total panel area. When the holes are evenly distributed on the total panel area, $S_u = 0$, and (2) is retrieved. Again, using the interstice length β leads to:

$$\sigma = \frac{n\pi a^2}{4(S_u + n(\beta + a)^2)}. \quad (6)$$

The interstice β is therefore always inversely proportional to the open area ratio *σ*. From this latter expression of *σ*, it is obvious that by varying the distance between the holes, it is possible to keep constant a fix value of the open area ratio and the diameter of holes by concentrating the holes in a part of the panel area. It would be preferable when studying the holes interaction effects to keep constant the open area ratio and the diameter of holes. The characteristics of the perforated panels tested in this work are given in table 1.

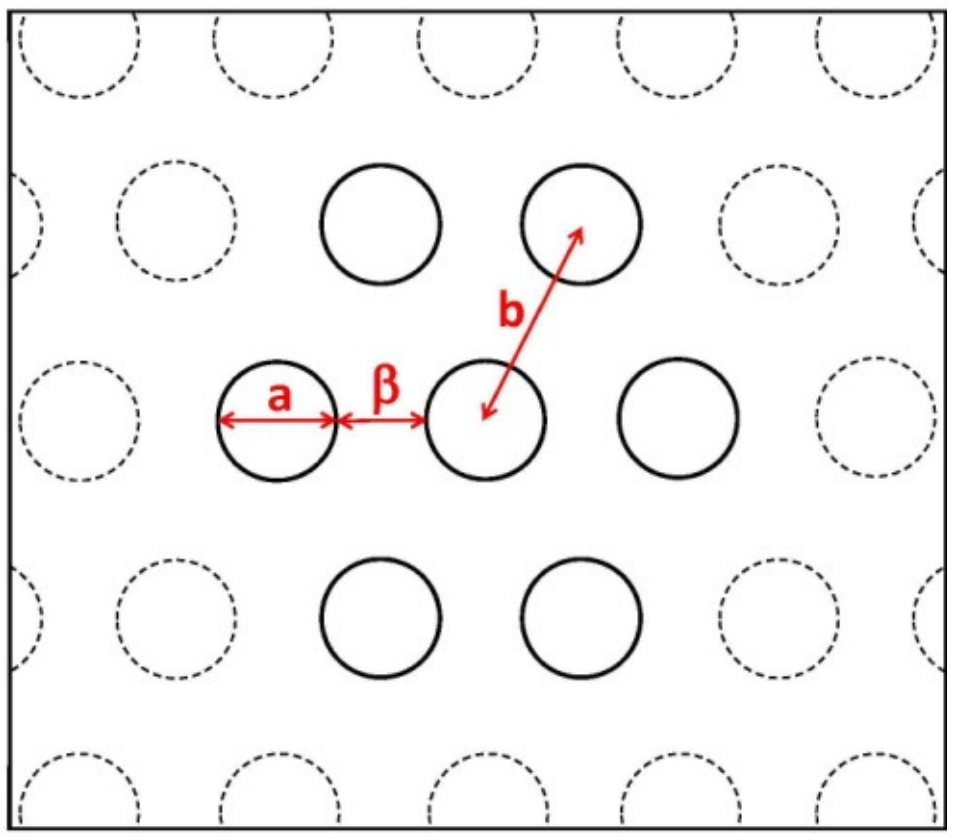

Fig 1: Schematic of the perforated panel and the nomenclature adopted. The dashed holes are the positions of the holes when the holes are evenly distributed on the total panel area.

## 1.1 The resulting impedance with holes interaction effect (HIE)

Considering the case of 2 apertures interacting, Ingard [4] found that the end correction is very dependent on the holes separation and that the interaction impedance can be determined from:

$$z_{1,2} = \frac{1}{U_2 S_2} \int_{S_2} p_{1,2} dS_2 \, , \tag{7}$$

where $p_{1,2}$ is the pressure exerted by hole 1 at hole 2, $U_2$ is the volume velocity through hole 2 and $S_2$ is the area of hole 2. Determining $z_{1,2}$ would lead to a new end-correction factor and therefore to a new specific impedance expression. Solving equation (7) for the case of infinitely thin plates, Fok's work [5, 6] provides an expression for an attached conductance and for a function of variable *a/b* expressed as :

$$\psi(\xi) = (1 + x_1\xi + x_2\xi^2 + x_3\xi^3 + x_4\xi^4 + x_5\xi^5 + x_6\xi^6 + x_7\xi^7 + x_8\xi^8)^{-1} \, , \tag{8}$$

where $\psi(\xi)$ is the so-called Fok function of variable a/b. With $x_1$=-1.4092; $x_2$=0; $x_3$=+0.33818; $x_4$=0; $x_5$=+0.06793; $x_6$=-0.02287; $x_7$=+0.03015; $x_8$=-0.01641; It's shown from Fig.6 of [7] that if *a/b* < 0.2 (which also corresponds to β > 4a), there is no appreciable interaction effect and that if *a/b* > 0.8 (which also corresponds to β < a/4), the attached mass is effectively zero.

To take into account the interaction effects of apertures, Rschevkin [6] proposed to modify the end-correction term of the perforated panel model with the Fok function. Due to flow of air through the holes which affects the air close to the inner and outer apertures, the thickness must be corrected with an additional term [1]. For circular cross-section holes, Rayleigh [9] proposed an end correction δ for the inner and outer aperture as follows:

$$\delta = \frac{8a}{3\pi} \tag{9}$$

Following Ingard's work, the acoustic specific impedance of a single aperture taking into account the inner and outer aperture is given by:

$$Z_1 = 4R_s\left(1 + \frac{h}{a}\right) + j\omega\rho_o(h + \delta) \tag{10}$$

where *h* is the panel thickness, $\rho_o$ is the density of air, ω is the pulsation and Rs is the surface resistance due to the viscous dissipation in the aperture and the surface of the panel given by:

$$R_s = \frac{1}{2}\sqrt{2\omega\rho_o\mu} \tag{11}$$

where μ is the coefficient of viscosity of air. Then, the total specific impedance $Z_p$ of the perforated panel with interacting holes is given by:

$$Z_P = 4R_s\left(1+\frac{h}{a}\right) + j\omega\rho_o(h + \frac{\delta}{\psi(\xi)}) \qquad (12)$$

where $\sigma$ is the open area ratio as described in the former section and $\psi(\xi)$ the Fok function. According to Ingard [4], under high sound intensities, assuming that the reactance slightly depends on the velocity, the resistance velocity-dependent expression is given by:

$$\frac{R_{nl}}{a} C_1(\frac{U}{100})^{C_2} \qquad (13)$$

where $R_{nl}$ is the nonlinear resistance, $U$ is the average incident velocity, $C_1$ and $C_2$ are constants that may depend on frequency. For a single frequency and using the Reynolds number, (13) can be transformed into:

$$R_{nl} = K_1 Re^{C_2} \qquad (14)$$

where $Re$ is the Reynolds number, $K_1$ and $C_2$ are the high sound levels coefficients determined experimentally. These coefficients lead to the best fitting of the nonlinear resistance curves. Their values for the samples used for measurements are given in table 2. The Reynolds number is expressed as:

$$Re = \frac{aU}{\nu} \qquad (15)$$

with ν the kinematic viscosity. If the perforated panel is backed by an air cavity, the resonant system formed has an impedance given by:

$$Z_S = Z_P + Z_{cav} \qquad (16)$$

where $Zcav$ is the air cavity impedance expressed as:

$$Z_{cav} = -jZ_o cot(k_o D_c) \qquad (17)$$

$Zo$ being the impedance of air, $k_o$ is the wave number and $D_c$ the air cavity depth. The reflection coefficient is obtained using the formula:

$$R = \frac{Z_S - Z_o}{Z_S + Z_o} \qquad (18)$$

And the absorption coefficient is deduced as:

$$\alpha = 1 - |R|^2 \qquad (19)$$

# 2. Experimental setup

## 2.1. The samples characteristics

All the measurements are performed on steel-made perforated panels of 2 mm thickness. Each panel sample has an external diameter of 100 mm. The mounting conditions of the samples inside the tube are closed to a clamped condition. Pictures of the samples used in the experiments are shown on Figs. 2, 3 and 4 and the samples characteristics are given in table 1.

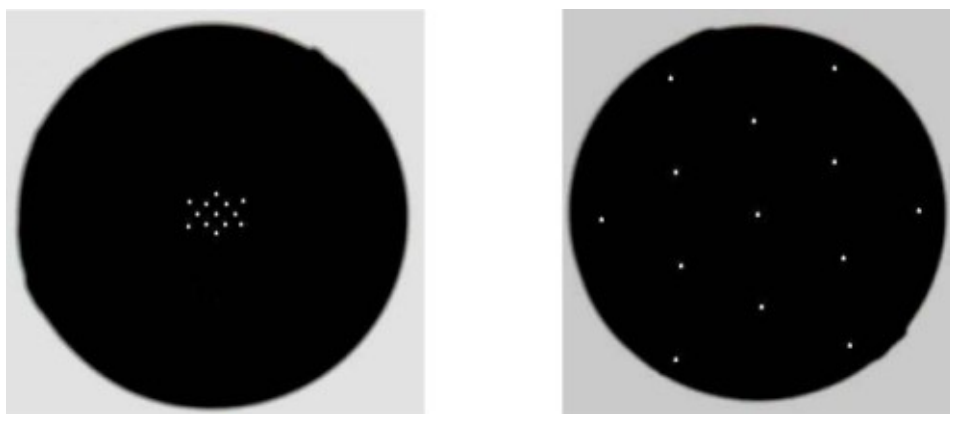

Fig 2: Samples pictures of panels 1a (left) and 1b (right)

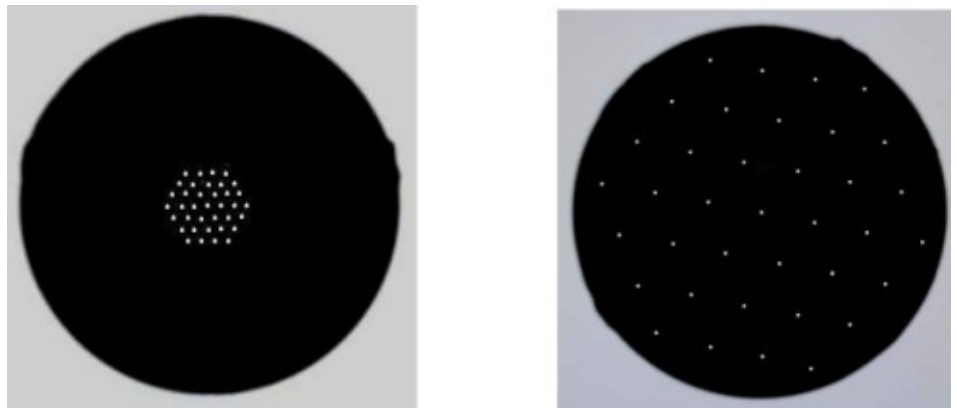

Fig 3: Samples pictures of panels 2a (left) and 2b (right)

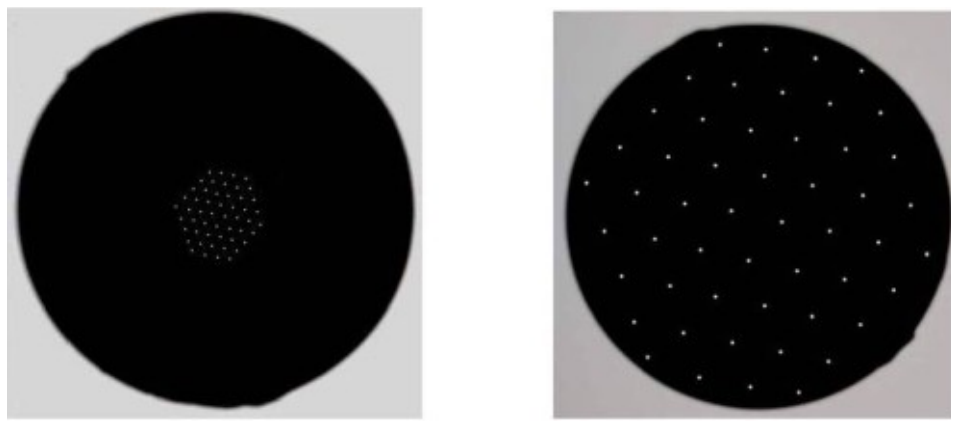

Fig 4: Samples pictures of panels 3a (left) and 3b (right)

Table 1: Perforated Panels characteristics.

| | a (mm) | σ (%) | β (mm) | |
|---|---|---|---|---|
| Panel 1a | 2.0 | 0.52 | 0.5 | Interaction |
| Panel 1b | 2.0 | 0.52 | 22.58 | No interaction |
| Panel 2a | 1.6 | 0.95 | 1.9 | Interaction |
| Panel 2b | 1.6 | 0.95 | 12.58 | No interaction |
| Panel 3a | 1.4 | 0.98 | 1.8 | Interaction |
| Panel 3b | 1.4 | 0.98 | 11.13 | No interaction |

## 2.1. The impedance tube and data acquisition

A schematic of the impedance tube used is shown in Fig. 5. It's a rigid circular plane-wave tube with a diameter of 100 mm (cut-off frequency of 1.7 KHz). At the left hand side, a compression driver JBL model 2450J is mounted as the source of excitation. This compression driver is powered by a professional power amplifier. A transition piece provides a continuity transition between the circular section of the compression driver and the circular cross section of the impedance tube. At the right hand side of the tube, a soundproof plunger is used as the rigid backing wall. The sealing for the plunger is ensured using rubber. By moving the plunger along the longitudinal axis of the tube, one is able to create an air cavity behind the sample. The sample is mounted between the speaker and the plunger. The whole tube has a thickness of 7mm to provide a sound-hard boundary condition. Three ¼'' microphones are used to perform the signal detection. The first two (micro 1 and 2 in Fig. 5) are used to calculate the surface impedance of the sample by the two microphones standing waves method described by Chung and Blaser [10]. The distance between these microphones is s = 50 mm. And the distance between micro 2 and the sample is about $l_2$ = 11 mm. The third microphone (reference micro in Fig. 5) acts as a reference microphone to get the level of pressure at the sample surface. An LMS System was used as the data acquisition unit. The measurements are performed taking a single sample panel with an air cavity of 20 mm behind and a rigid wall. A first excitation is done with a periodic random noise signal in order to have a general view of the absorption coefficient curve and locate the viscous peak position. The result is used to get the resonant frequency and to perform a second excitation (sine excitation) at the resonance frequency. The amplitude of the source is adjusted such that the sound pressure level measured by the reference microphone is set at the level of interest. The SPL (Sound Pressure Level) is varied from 100-150 dB at the face of the sample monitored using the reference microphone. It is insured that for these levels, there is neither saturation nor great bifurcation occurring on the signal for the desired frequencies. The sound pressure reference used is 20 µPa. A calibration phase of the microphones is first of all done for the separate phase and magnitude calibration for the two microphones. The calibration transfer function $H_{cal}$ between the two microphones is saved. This saved calibration transfer function is combined with the measured frequency response $H_{12meas}$ to calculate the true frequency response H as :

$$H = \frac{H_{12meas}}{H_{cal}} \tag{20}$$

The values of the velocity shown in the experimental results are the viscous peak corresponding particle velocity (see Dalmont [11]) given by:

$$u = j\frac{P_1}{Z_o}\frac{H\cos(k_o l_1)-\cos(k_o l_2)}{\sin(k_o s)} \tag{21}$$

where $l_1$=s+$l_2$ (Fig. 5), $p_1$ is the pressure on microphone 1, $l_1$ (resp. $l_2$) is the distance from microphone 1 (resp. 2) to the panel sample.

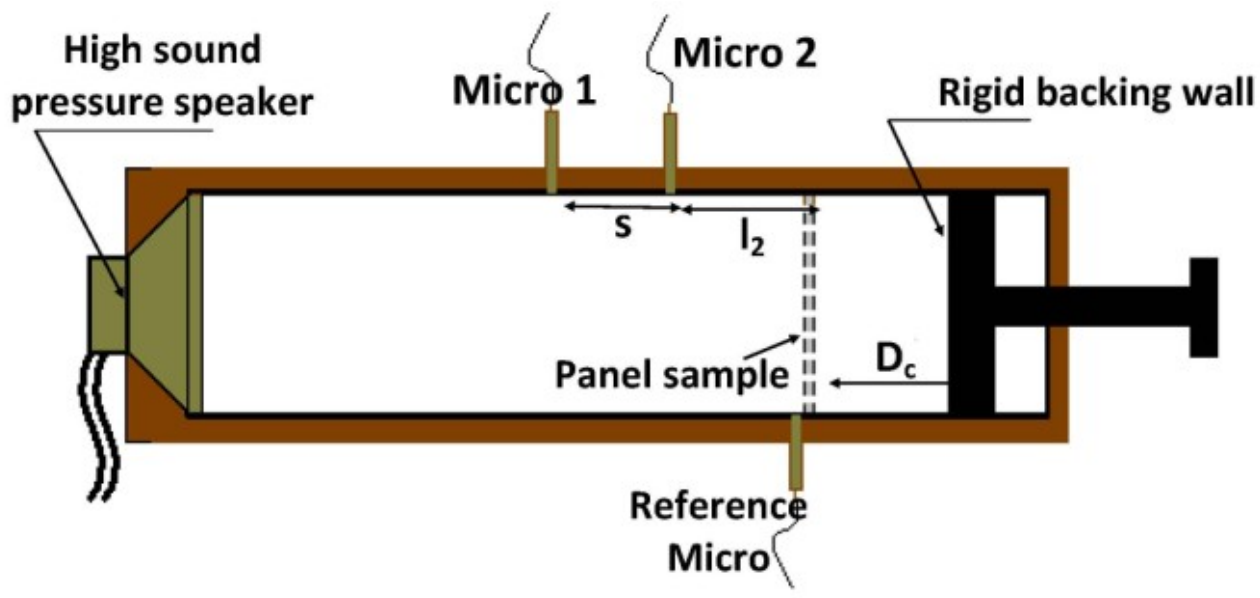

Fig 5: Schematic of the impedance tube used for the measurements.

Table 2: The high sound intensities constants of the samples

| | $K_1$ | $C_2$ |
|---|---|---|
| Panel 1a | 1.90 | 0.49 |
| Panel 1b | 4.82 | 0.32 |
| Panel 2a | 0.72 | 0.62 |
| Panel 2b | 0.50 | 0.77 |
| Panel 3a | 0.63 | 0.61 |
| Panel 3b | 1.06 | 0.52 |

# 3. Results and discussion

In this section, measurement results of perforated panels backed by an air cavity are presented and confronted to the simulated model described above in the theory section. Unless specified differently, the air cavity depth behind the sample is of 20 mm. In the experimental figures, empty symbols correspond to cases of no holes interaction effect and the filled symbols correspond to the cases with holes interaction effect.

## 3.1. Hole interaction effect on the nonlinear resistance

Fig. 6 shows the experimental and simulated results of the surface resistance as a function of the Reynolds number in front of the panel. It is seen that the HIE causes a decrease of the perforated plate resistance with the increase of the Reynolds number. For high sound pressure levels, where the nonlinear resistance is usually greater than the medium impedance $Z_o$, this decrease of the nonlinear resistance may improve the absorption coefficient. However, since the decrease of the nonlinear resistance due to the HIE is not constant, the slope of the curve changes from the case without HIE to the case with HIE. It's known that for high sound pressure levels, the dependency of the resistance with the velocity is linear [4,7]. The HIE changes this slope and therefore, to accurately predict the nonlinear behavior in the case with HIE, the expression of this slope should take into account (directly or indirectly) the interstice of the perforated panel.

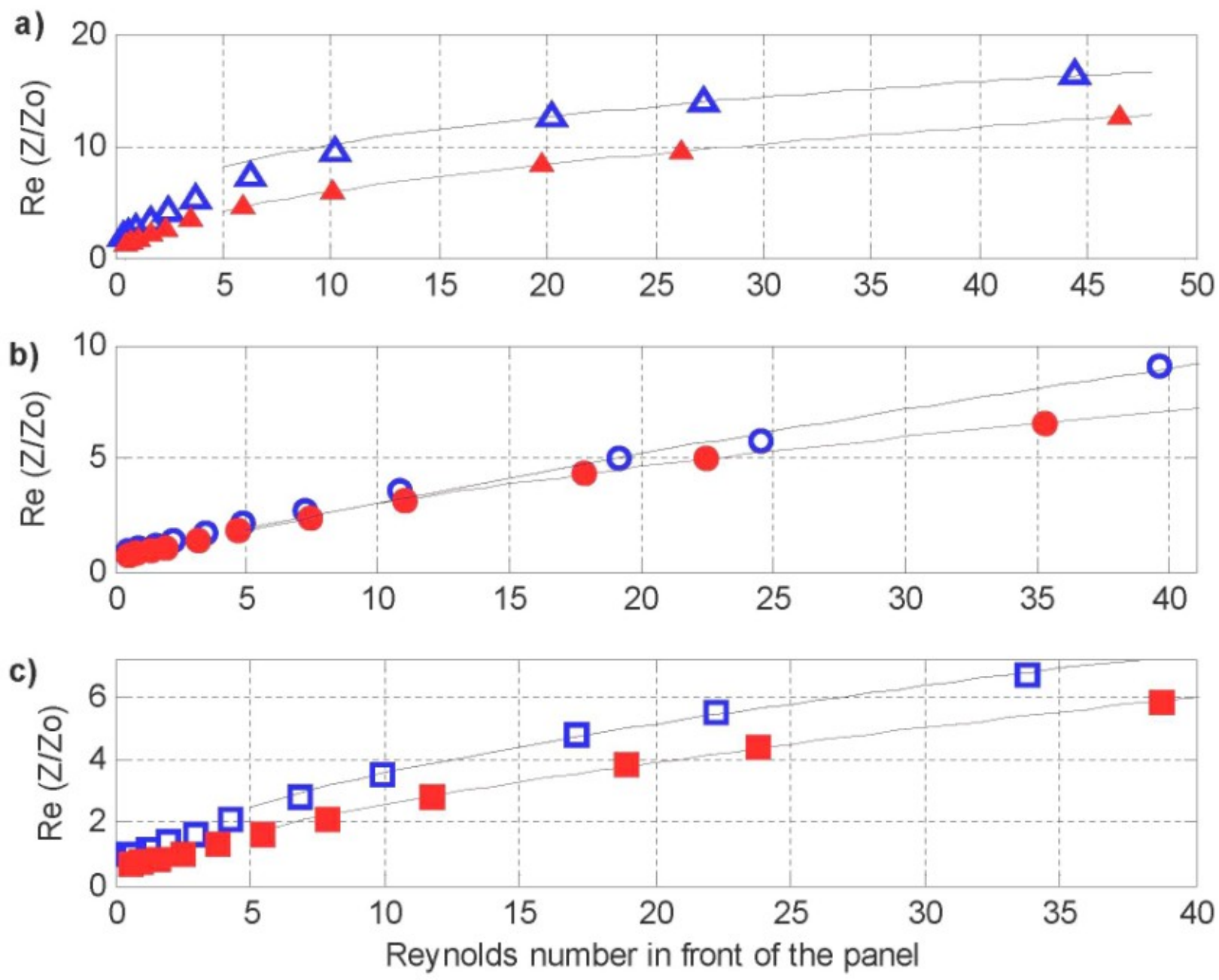

Fig 6: Experimental and simulated results of the normalized surface resistance as a function of the Reynolds number in front of the panel. Dotted lines and Filled symbols=With HIE; Dashed lines and Empty symbols=Without HIE. a) panel 1a (368 Hz) & 1b (406 Hz); b) panel 2a (528Hz) & 2b (666 Hz); c) panel 3a (600 Hz) & 3b (646 Hz). Air cavity of 20 mm.

## 3.2. Hole interaction effect on the nonlinear reactance

Fig. 7 shows the experimental results of the surface reactance as a function of the Reynolds number in front of the panel. The reactance is known to be linked to the added mass effect. In this figure, it is observed that with the increase of Reynolds number, the reactance decreases in a first phase before tending in taking a relatively constant value. This is a typical nonlinear behavior of the reactance of thin panels whenever turbulence is reached. This behavior, which was also already noticed in [4], is caused by a transfer of part of the kinetic energy in the sound field around the aperture into turbulent motion that breaks away from the aperture. With the HIE, this behavior is still observed for all the samples. In Fig. 7a) and Fig. 7b), with the HIE the reactance is throughout negative and smaller than the reactance without HIE. Meanwhile, in Fig. 7c), the reactance with HIE is in a first phase positive and greater than the reactance without HIE and in a second phase smaller than the reactance without HIE. In Fig. 7, the simulation results are not presented because of the theoretical assumption that is made that the reactance does not depend on the incident velocity. The dependency is so slight that it can globally be considered as constant.

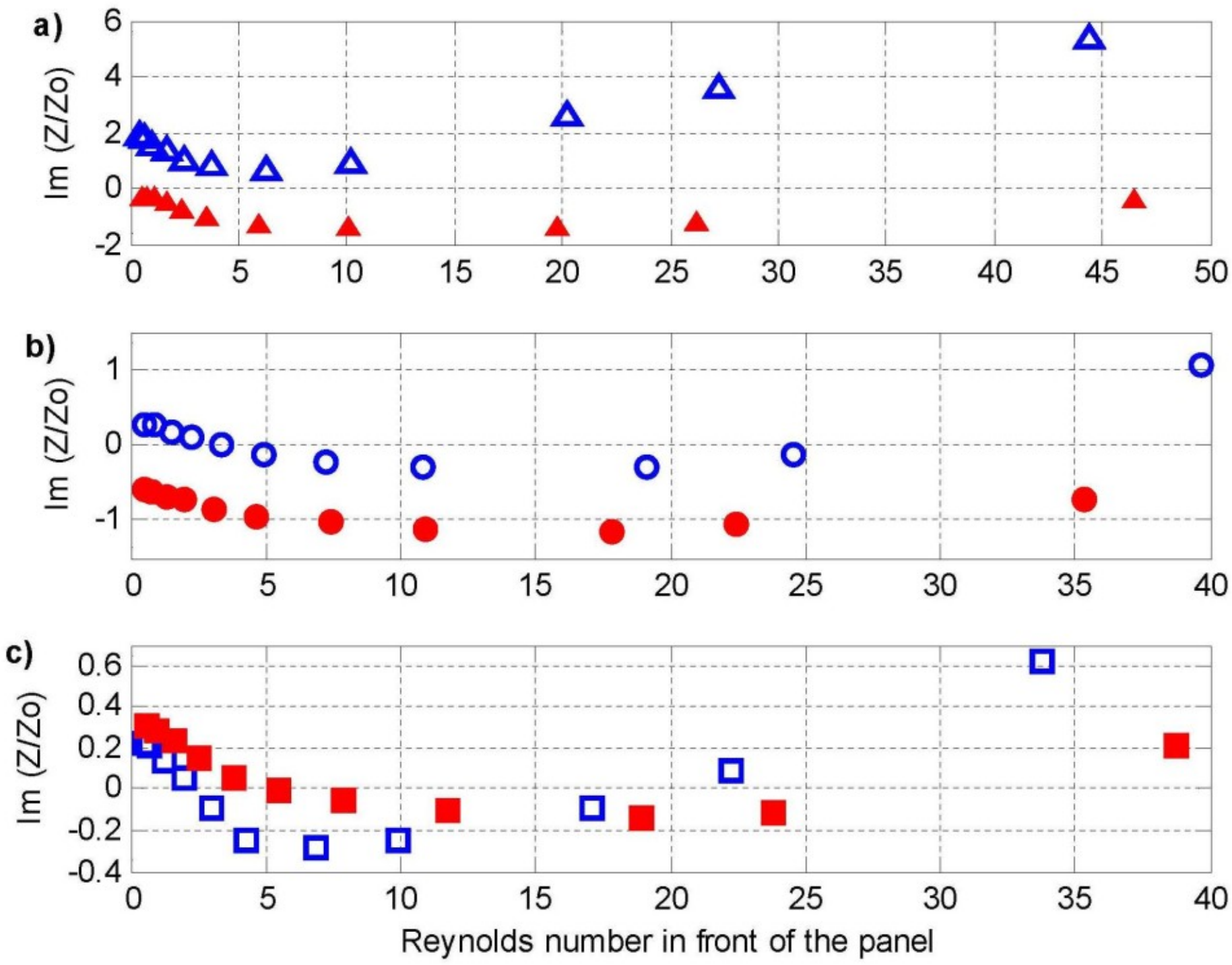

Fig 7: Experimental result of the normalized surface reactance as a function of the Reynolds number in front of the panel. Filled symbols: With HIE; Empty symbols: Without HIE. a) panel 1a (368 Hz) & 1b (406 Hz); b) panel 2a (528 Hz) & 2b (666 Hz); c) panel 3a (600 Hz) & 3b (646Hz). Air cavity of 20 mm.

Fig. 8 shows the experimental results of the surface reactance discrepancies (with and without HIE) as a function of the Reynolds number for panel 1a and 1b (**), panel 2a and 2b (++), panel 3a and 3b (xx). Except for panels 1a and 1b, one can notice that the reactance discrepancies (with and without HIE) are globally almost constant and low throughout the whole range of Reynolds number. This reveals the fact that even with holes interaction effects the assumption that for high sound intensities the reactance is slightly dependent of velocity may be acceptable only if the interstice β is much greater than the quarter of holes diameter. For panel 1a, for which the interstice β is equal to the quarter of the diameter (β=a/4), Fok function is very great and thus provides a zero mass attached to the reactance with HIE. This may be a reason why the constant reactance remark does not apply in this case.

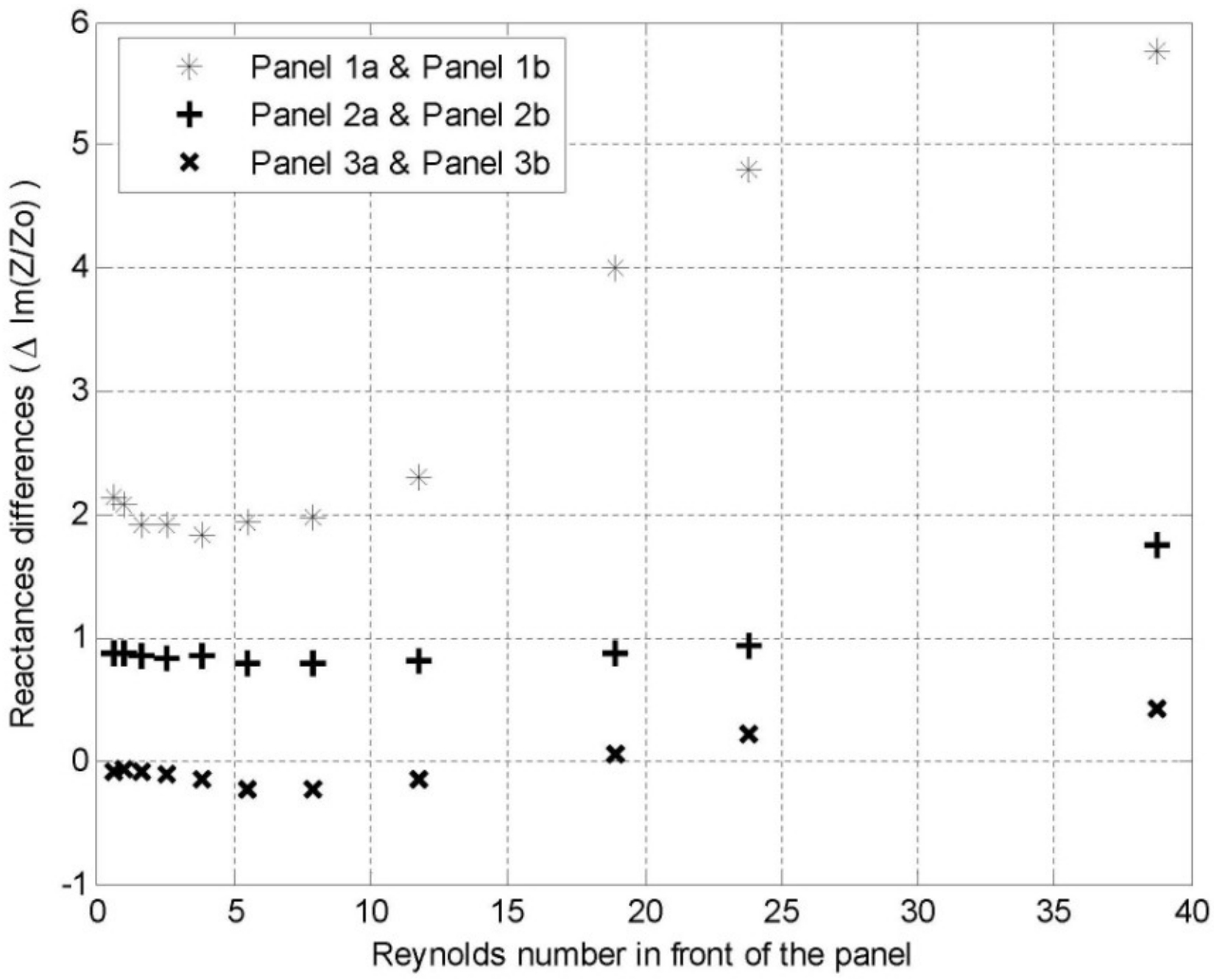

Fig 8: Experimental result of the normalized reactance differences (with and without HIE) as a function of the Reynolds number in front of the panel. Air cavity of 20 mm.

## 3.2. Hole interaction effect on the absorption coefficient

Fig. 9 shows the experimental and simulated results of the absorption coefficient as a function of the Reynolds number in front of the panel. The predictions are higher than the measurements. Yet the tendency is observed. It is observed that HIE does not affect much the tendency that with increase of intensities, the absorption coefficient increases to a maximum value before decreasing. The point (critical point) after which this decrease starts can be considered. This critical point will not always be noticeable for some panels. On Fig. 9a), though the fact that the critical point does not appear for the case with HIE, better absorption is achieved with HIE throughout the range of the Reynolds numbers at low and high intensities. This probably comes from the fact that since the interstice of panel 1a is equal to the quarter of the hole diameter, it creates a smaller (compared to the case with interactions) but enough kinetic energy around the aperture so as to optimize the absorption coefficient. The optimized absorption coefficient is known to be achieved when the reactance is equal to zero and the resistance equal to the medium resistance. On Fig. 9b), the critical point is observable and located at almost the same Reynolds number for both cases with and without HIE. Up to a certain Reynolds number of 10.84, better absorption is achieved without HIE. Beyond this Reynolds number of 10.84, the absorption coefficient with HIE is also interesting. On Fig. 9c), the critical point is moved to a higher Reynolds number for the case with HIE. Therefore, this reveals the fact that the critical point is related not only to the shape of the aperture but also to the interstices of the panel. Beyond this critical point and throughout the whole range of Reynolds number the best absorption is provided by panel 3a which corresponds to the case with HIE.

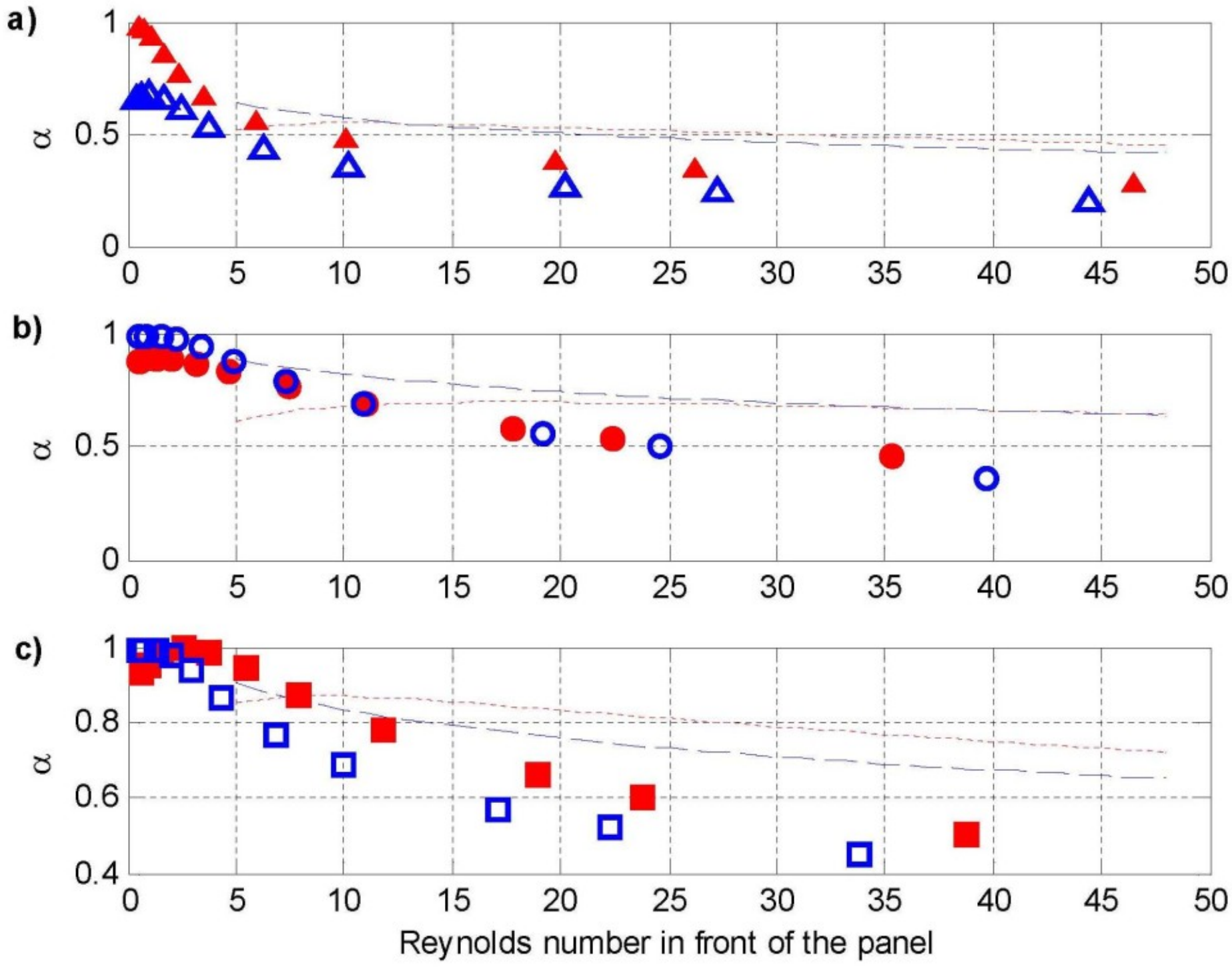

Fig 9: Experimental and simulated results of the absorption coe_cient (with and without HIE) as a function of the Reynolds number in front of the panel. Dotted lines and Filled symbols=With HIE; Dashed lines and Empty symbols=Without HIE. a) panel 1a (368 Hz) & 1b (406 Hz); b) panel 2a (528 Hz) & 2b (666 Hz); c) panel 3a (600 Hz) & 3b (646 Hz). Air cavity of 20 mm.

# Conclusion

An experimental investigation of holes interaction effects of air-cavity-backed perforated panels under high sound pressure levels was carried out in this paper. It was demonstrated experimentally that even with holes interaction effects, the nonlinear reactance is still slightly dependent of the incident velocity assuming that the interstices (distance between two consecutive holes edges) is greater than the quarter of holes diameter. It was also noticed that the interaction effect can improve the absorption coefficient of the perforated panel system for low and high sound levels. One can therefore use this interaction effects to increase the critical point after which the absorption coefficient starts to decrease with the increase of sound intensity. With this shift of the critical point, a certain compromise may be made between low and high intensity absorptions. Nevertheless, this critical point definitely depends also of the aperture exit (shape and neighboring). This study was done only for the case of normal incident wave, and it would be interesting looking into these interaction effects when submitted to bias or grazing flow under high levels of excitation.